\documentclass[10pt]{iopart}

\usepackage{graphicx}
\usepackage{exscale}
\usepackage{iopams}
\usepackage{subfigure}
\usepackage{xspace}
\usepackage{citesort}

\newcommand{\text}[1]{\mathrm{#1}}

\renewcommand{\Im}[0]{\mathrm{Im}\,}
\renewcommand{\Re}[0]{\mathrm{Re}\,}

\newcommand{\ie}[0]{i.e.\@\xspace}
\newcommand{\eg}[0]{e.g.\@\xspace}

\newcommand{\om}[0]{\omega}

\newcommand{\Ep}{E_\mathrm{P}}
\newcommand{\rD}[0]{\mathrm{D}}

\newcommand{\nag}{\phantom{\dag}}

\newcommand{\on}{\hat{n}}

\newcommand{\Ef}{E_\text{F}}

\newcommand{\las}[0]{\langle}
\newcommand{\ras}[0]{\rangle}
\newcommand{\la}[0]{\left\las}
\newcommand{\ra}[0]{\right\ras}
\newcommand{\ket}[1]{\left|#1\ra}
\newcommand{\bra}[1]{\la#1\right|}

\begin{document}


\title{Carrier-density effects in many-polaron systems}

\author{M Hohenadler,$^1$ G Hager,$^2$ G Wellein,$^2$ and H Fehske$^3$}

\address{$^1$%
  Theory of Condensed Matter, Cavendish Laboratory, University of Cambridge, United Kingdom
}
\address{$^2$%
  Computing Centre, University of Erlangen, Germany
}
\address{$^3$%
  Institute of Physics, Ernst-Moritz-Arndt University Greifswald, Germany
}

\ead{\mailto{mh507@cam.ac.uk}}

\begin{abstract}
  Many-polaron systems with finite charge-carrier density are often encountered
  experimentally. However, until recently, no satisfactory 
  theoretical description of these systems was available 
  even in the framework of simple models such
  as the one-dimensional spinless Holstein model considered here. In this
  work, previous results obtained using numerical as well as analytical
  approaches are reviewed from a unified perspective, focussing on spectral
  properties which reveal the nature of the quasiparticles in the system. In
  the adiabatic regime and for intermediate electron-phonon coupling, a
  carrier-density driven crossover from a polaronic to a rather metallic
  system takes place. Further insight into the effects due to changes in
  density is gained by calculating the phonon spectral function, and the
  fermion-fermion and fermion-lattice correlation functions. Finally, we
  provide strong evidence against the possibility of phase separation.
\end{abstract}



%
%
%
\section{Introduction}\label{sec:intro}
%
%
%

The concept of a polaron as a charge carrier bound to a self-created lattice
distortion (polarisation) as a result of electron-phonon (EP) interaction has
been introduced long ago by Landau \cite{Landau33}. In recent decades,
experimental work on a variety of materials has revealed the existence of
such quasiparticles in many cases. Of particular interest in this context are
colossal-magnetoresistive manganites, in which there is ample experimental
evidence for the polaronic character of the charge carriers \cite{David_AiP}.
More recently, a lot of work on polaron physics has been driven by
technological realization of nanostructures such as quantum wells or dots, in
which the confinement of carriers usually enhances lattice effects (see
\cite{HoFe06} and references therein).

Until recently, the bulk of theoretical work on polaronic systems was
concerned with the zero-density case (one or two electrons) for which, within
the framework of the one-dimensional (1D) Holstein model considered here, it
is now well understood that a crossover takes place from a large polaron
(extending over more than one lattice site) to a small polaron (with the
lattice distortion essentially being localised to the same lattice site as
occupied by the electron) upon increasing the EP interaction strength
\cite{dRLa82,AlMo95} (for a review of spectral properties see
\cite{FeAlHoWe06,FeTr2006}). The critical coupling for the crossover sensitively
depends on the ratio of the phonon frequency and the electronic hopping
integral. Despite a noticeable increase in effective mass in the
strong-coupling regime, in a strict sense, 
polarons remain itinerant quasiparticles at zero
temperature.  Quantum phonon fluctuations---not to be 
neglected in any reliable calculation---strongly affect the 
transport properties, especially in low-dimensional systems.

Although a detailed understanding of the process of polaron formation in the
low-density limit was a necessary first step, real materials are usually
characterised by finite charge-carrier densities, so that 
residual interaction between individual polarons (due to overlap of the
phonon clouds) becomes important.  For parameters relevant to experiment, a
rigorous treatment of density effects
\cite{Al92,CaGrSt99,HoNevdLWeLoFe04,HoWeAlFe05,WeBiHoScFe05} gives rise to
interesting new results~\cite{HoNevdLWeLoFe04,HoWeAlFe05,WeBiHoScFe05},  
substantially different from those for a gas of weakly or 
non-interacting polarons realised, \eg, in 
the non-adiabatic regime. These discrepancies may also 
hint at an explanation of the
difficulties and inconsistencies arising when fitting experimental data on,
\eg, the manganites by using results valid for independent polarons or weak
coupling \cite{David_AiP,HaMaDeLoKo04}.

In this paper, we review and extend recent results on the so-called
many-polaron problem in the framework of the 1D spinless Holstein model. To
this end, we report on numerical data from exact diagonalisation, cluster
perturbation theory and the density-matrix renormalisation group, as well as
analytical self-energy calculations, to study photoemission and phonon
spectra, optical conductivity as well as static correlation functions. The
numerical methods fully take into account quantum phonon effects, and are
capable of describing the most important regime of intermediate EP coupling
at finite charge-carrier densities.

The paper is organised as follows. The model is introduced in
section~\ref{sec:model}, and a quick overview of the methods employed and a
definition of observables is given in section~\ref{sec:methods}.
Section~\ref{sec:results} is devoted to a discussion of previous and new
results on the density dependence of the physics. Finally, we conclude in
section~\ref{sec:conclusions}.

%
%
\section{Model}\label{sec:model}
%
%
%

As we shall see in section~\ref{sec:results}, the density dependence of, in
particular, the photoemission spectra is quite complicated. Therefore, it is
necessary to begin with a simple yet physically reasonable model for a
many-polaron system. In this way, the density effects may be understood
without additional complications due to, for example, Mott-Hubbard physics
occurring in the spinful case \cite{FeWeHaWeBi03}. Hence we shall consider
the Hamiltonian
\begin{equation}\label{eq:model:H}
  H
  =
  -t \sum_{\las i,j\ras} c^\dag_i c^{\nag}_j
  +\om_0\sum_i b^\dag_i b^{\nag}_i
  -g\om_0\sum_i \on_i (b^\dag_i + b^{\nag}_i)
  \,.
\end{equation}
Here $c^\dag_i$ ($c^{\nag}_i$) creates (annihilates) a spinless fermion at
lattice site $i$, $b^\dag_i$ ($b^{\nag}_i$) creates (annihilates) a phonon at
site $i$, and $\on_i=c^\dag_i c^{\nag}_i$.  The first term of
Hamiltonian~(\ref{eq:model:H}) describes the hopping of spinless fermions
between neighbouring sites $\las i,j\ras$ on a 1D lattice. 
The lattice constant is taken to be unity.
The second term accounts for the elastic and kinetic energy of 
the lattice ($\hbar=1$) and, finally, the last term
constitutes a local coupling of the lattice displacement to the fermion
occupation number which can be zero or one.

The model parameters are the hopping integral $t$, the phonon 
frequency $\om_0$, and the coupling constant $g$. 
The relation of the atomic-limit
polaron binding energy to the latter reads $\Ep=g^2\om_0$. Introducing the
dimensionless quantities $\lambda = \Ep/2t$ and $\gamma=\om_0/t$ we are left
with two independent parameter ratios. 

%
%
\section{Methods}\label{sec:methods}
%
%
%

In order to achieve a thorough understanding of the carrier-density
dependence of the Holstein model, it is beneficial to make use of
both numerical and analytical methods. Whereas large-scale numerical
simulations yield exact results for finite systems, approximate
analytical calculations yield additional insight into the problem. The
methods to be used here have been described in detail elsewhere
\cite{HoNevdLWeLoFe04,WeWeAlScFe05,JeFe06,LoHoFe06}.

\subsection{Numerical methods}

In this paper we use exact diagonalisation (ED) 
in combination with cluster perturbation theory 
(CPT) \cite{SePePL00,HoAivdL03} and the kernel polynomial
method (KPM) \cite{WeWeAlScFe05}, as well as the density-matrix 
renormalisation group (DMRG) (for adaption of these techniques 
to coupled EP systems see~\cite{JeFe06} and references therein). 
In all cases we employ parallel codes to obtain
reliable results for the complex many-body problem under consideration.
Furthermore, for most calculations, the homogeneous $q=0$ lattice distortion
has been treated separately to reduce the phonon Hilbert space
\cite{SyHuBeWeFe04}.  All results of
this work are for zero temperature. Note that we have in the 
past also applied a (finite-temperature) quantum Monte Carlo (QMC) algorithm 
to the present problem~\cite{HoNevdLWeLoFe04}.
Although the results are consistent with the findings from other
approaches, it turns out that the subtle details of the crossover are
difficult to see in the calculated single-particle spectra due to temperature
effects and the use of the maximum entropy method.

The main observable of interest here is the one-particle spectral function,
which provides us with valuable information about the character of the
quasiparticles in the system, as well as about the existence of excitation
gaps, etc. The transition amplitude for removing ($-$) (adding ($+$)) a free
electron---corresponding to (inverse) photoemission---is determined by the
imaginary part of the retarded one-particle Green function
\begin{eqnarray}\label{eq:Gkw}
  G^\pm(k,\omega)
  =
  \lim_{\eta\to 0}\ \bra{\psi_0} c_k^\mp 
  \frac{1}{\omega +\mathrm{i}\eta -H} c_k^\pm \ket{\psi_0}
  \,,
\end{eqnarray}
and hence by the wavevector-resolved spectral function
\begin{eqnarray}\label{eq:akw}
  A^{\pm}(k,\omega) 
  =
  -\frac{1}{\pi}\mathrm{Im}\,G^\pm(k,\omega)
  \,.
\end{eqnarray}
Here $c_{k}^{+}=c_{k}^{\dagger}$, $c_{k}^{-}=c_{k}^{}$, and $\ket{\psi_0}$
denotes the ground state of Hamiltonian~(\ref{eq:model:H}). 

The spectral functions~(\ref{eq:akw}) can be calculated exactly on small
clusters with $N$ lattice sites (and hence for a finite number of
wavevectors) with periodic boundary conditions using the KPM. 
Alternatively, using CPT, we may calculate
the real-space cluster Green function $G^c_{ij}(\omega)$ of a
$N_\text{c}$-site cluster with open boundary conditions for all
non-equivalent combinations of $i,j=1,\dots,N_\text{c}$, from which an
approximation for the spectrum of the infinite lattice can be obtained by
treating the inter-cluster hopping in first-order strong-coupling
perturbation theory \cite{SePePL00}. Then the partial
densities of states (DOSs) are obtained from 
\begin{equation}\label{eq:dos}
  \rho^\pm(\om)
  = 
  N_\text{(c)}^{-1} \sum_k A^\pm(k,\om)
  \,.
\end{equation}

The phonon spectral function 
\begin{equation}\label{eq:bqw_cpt}
  B(q,\om)= -\frac{1}{\pi}\Im D^\mathrm{R}(q,\om)
\end{equation} 
is calculated by using a cluster approximation
\cite{FeAlHoWe06,FeTr2006,HoWeBiAlFe06,LoHoAlFe06} for the phonon Green function
\begin{equation}\label{eq:phononSF}
  D^\mathrm{R}(q,\om>0) 
  = 
  \lim_{\eta\to 0^+}\ 
  \las \psi_0 | x_{q} 
  \frac{1}{\om + \rmi\eta - H}
  x_{-q} |\psi_0\ras
  \,,
\end{equation}
with the phonon coordinates $x_{q}=N_\text{c}^{-1/2}\sum_j x_j
e^{-\rmi qj}$,  $x_i=b^\dag_i+b^{\nag}_i$, and $B(\pm q,\om)=B(q,\om)$.

We shall also present KPM results for the linear optical response
$\Re\,\sigma(\omega) = \mathcal{D}\delta(\omega)+\sigma^\text{reg}(\omega)$
to a longitudinal electric field, with the regular part
\begin{equation}\label{eq:oc}
  \sigma^\text{reg}(\omega)
  =
  \frac{\pi}{N} \sum_{m>0} 
  \frac{|\bra{\psi_0}\hat{\jmath}\ket{\psi_m}|^2}{E_m-E_0}\, 
  \delta[\omega - (E_m-E_0)]
  \,.
\end{equation}
Here $E_m$ is the energy of the $m$-th eigenstate $\ket{\psi_m}$ of the
Hamiltonian~(\ref{eq:model:H}) defined on an $N$-site lattice with 
periodic boundary conditions, and
the current operator $\hat{\jmath}=\rmi\, e t \sum_i (c^\dag_i c^{\nag}_{i+1}
- c^{\dag}_{i+1}c^{\nag}_i)$.

Finally, we consider the fermion-boson and fermion-fermion correlation functions
\begin{equation}\label{eq:Cep}
  C_\text{ep}(r)
  =
  \sum_i
  \las (\on_i-n) x_{i+r}\ras
\end{equation}
and
\begin{equation}\label{eq:Cee}
  C_\text{ee}(r)
  =
  \sum_i
  \las (\on_i-n)(\on_{i+r}-n)\ras
  \,,
\end{equation}
respectively. 
Here $r=0,\dots,N-1$, and we have introduced the notation
$n=N_\text{e}/N$ for the carrier density, where $N_\text{e}$ denotes
the number of fermions in the ground state.

\subsection{Analytical method}

On the analytical side, we shall present recent results from second-order
self-energy calculations valid in the weak- and strong-coupling regime,
respectively \cite{LoHoFe06}.  This approach allows us to calculate the
coherent (``c'', infinite lifetime) and incoherent (``ic'', finite
lifetime) contributions to the one-particle (polaron) spectral function
given by \cite{LoHoFe06}
\begin{equation}
  A_\text{(p)}(k,\om)
  =
  A^\text{c}_\text{(p)}(k,\om)\left.\right|_{|\om|<\om_0}
  +
  A^\text{ic}_\text{(p)}(k,\om)\left.\right|_{|\om|\geq\om_0}
  \,,
\end{equation}
where all energies are measured relative to the Fermi energy $\Ef$.

%
%
\section{Results}\label{sec:results}
%
%
%

From previous work \cite{CaGrSt99,HoNevdLWeLoFe04,LoHoFe06} (see also
references in \cite{LoHoFe06}), the basic effects of variations of the
carrier density in the framework of the Hamiltonian~(\ref{eq:model:H}) are
known. It turns out that the physics is particularly simple in the
non-adiabatic regime $\gamma\gg1$, where the quasiparticles are small
polarons at all densities at intermediate ($g\simeq 1$) 
and strong ($g\gg 1$) EP coupling~\cite{CaGrSt99,HoNevdLWeLoFe04,LoHoFe06}. 
In contrast, in the experimentally
often realised adiabatic regime $\gamma\ll 1$, recent studies have revealed
important new effects to be discussed in detail below.  Apart from providing
a unified discussion of previous work, we present new results to extend the
present knowledge about the 1D spinless Holstein model.

We shall begin with the limiting cases of weak and strong EP coupling,
respectively, for which the calculated spectral functions are rather simple.
Due to the shortcomings of the QMC method used previously, we apply ED and
CPT with significantly better energy resolution.  Following previous work, we
set $\gamma=0.4$.

The one-electron ground state of the Holstein model is a 
(weakly-dressed electron) large polaron at (very) weak EP coupling, and a
small polaron at strong coupling. For $\gamma<1$, the crossover between
these two regimes takes place at about $\lambda=1$ \cite{FeAlHoWe06,FeTr2006}.

\begin{figure}
  \begin{center}
  \includegraphics[width=0.495\textwidth]{cpt_wc_n0.4.eps}
  \includegraphics[width=0.495\textwidth]{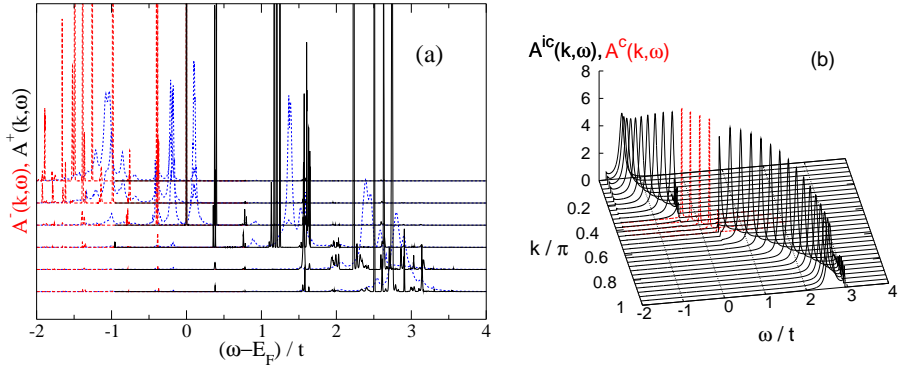}
  \end{center}
  \caption{\label{fig:cpt_an_akw_wc}%
    (colour online) (a) Exact diagonalisation results ($N=10$) for
    the single-particle spectral functions $A^-(k,\omega)$ ($broken$, red)
    and $A^+(k,\omega)$ ($\full$, black). Also shown are CPT results for
    $N_\text{c}=10$ ($\dotted$, blue).  (b) Analytical results for the
    coherent ($A^\text{c}(k,\omega)$, $\broken$, red) and incoherent
    ($A^\text{ic}(k,\omega)$, $\full$, black) spectral function from the
    weak-coupling approximation.  Here $\gamma=0.4$, $\lambda=0.1$, and
    $n=0.4$.}
\end{figure}

\subsection{Weak coupling}\label{sec:wc}

For $\lambda\ll1$, the EP interaction slightly renormalises the charge
carriers. Even for large densities $n$, the spectrum does not change
qualitatively \cite{HoNevdLWeLoFe04}. Figure~\ref{fig:cpt_an_akw_wc}(a) shows
ED and CPT spectra for $\lambda=0.1$ and $n=0.4$. In accordance with QMC
calculations \cite{HoNevdLWeLoFe04}, the ED results reveal a free-electron
like main band (consisting of several discrete peaks on a finite cluster, cf
CPT and figure~\ref{fig:cpt_an_akw_wc}(b)) running from -1 to 3 (\ie, having
almost the bare bandwidth $4t$). In the vicinity of the Fermi level $\Ef$, the
spectrum is dominated by large coherent peaks---small peaks are visible at
energies separated from $\Ef$ by multiples of $\om_0$---whereas significant
incoherent contributions exist for small or large $k$. Hence, already for
$\lambda=0.1$, non-trivial effects due to EP coupling are visible.

Figure~\ref{fig:cpt_an_akw_wc}(a) also includes CPT results for
$N_\text{c}=10$. However, although reliable results have been obtained with
this method in the intermediate coupling regime (see below), CPT fails for
the parameters used here. Despite the existence of excitations related to the
peaks of the exact results, we find spurious additional peaks having significant
spectral weight which are due to the open boundary conditions and the
perturbative treatment of the hopping term (see also \cite{HoWeBiAlFe06}).  
For CPT to yield reliable results at weak coupling, even larger
clusters would be required to capture the important non-local correlations.
Alternatively, the hopping to neighbouring clusters may be treated as a
variational parameter \cite{PoAiDa03}.

Naturally the small value of $\lambda$ motivates the use of 
perturbation theory (in the coupling term). 
Such calculations, also valid at finite charge-carrier
density, have been carried out in \cite{LoHoFe06}, and results from the
weak-coupling approximation are presented in
figure~\ref{fig:cpt_an_akw_wc}(b).

The overall agreement with ED
(figure~\ref{fig:cpt_an_akw_wc}(a)) is surprisingly good. The width of the
main band fits well that of the numerical spectrum, and even the
low-(high-)energy phonon features at small (large) wavevectors are
reproduced. Besides, similar to ED, the analytical approach predicts coherent
excitations with infinite lifetime for energies $|\om|<\om_0$, whereas the
CPT spectrum reveals a multi-peak structure and damping even close to $\Ef$.

\begin{figure}
  \begin{center}
  \includegraphics[width=0.495\textwidth]{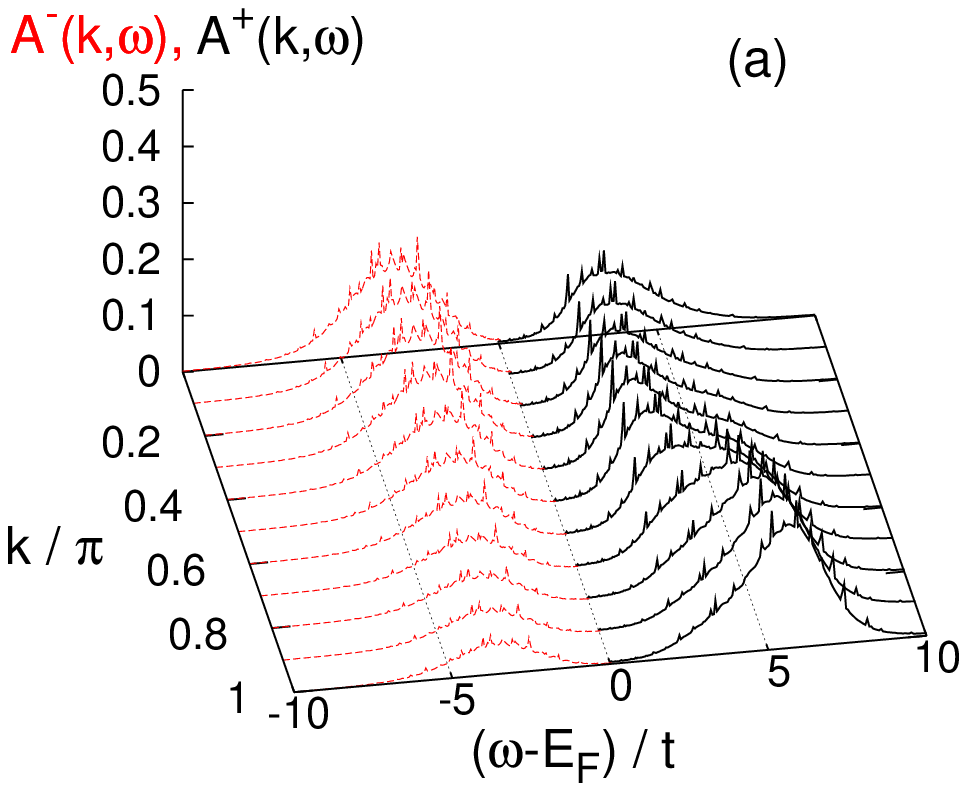}
  \includegraphics[width=0.495\textwidth]{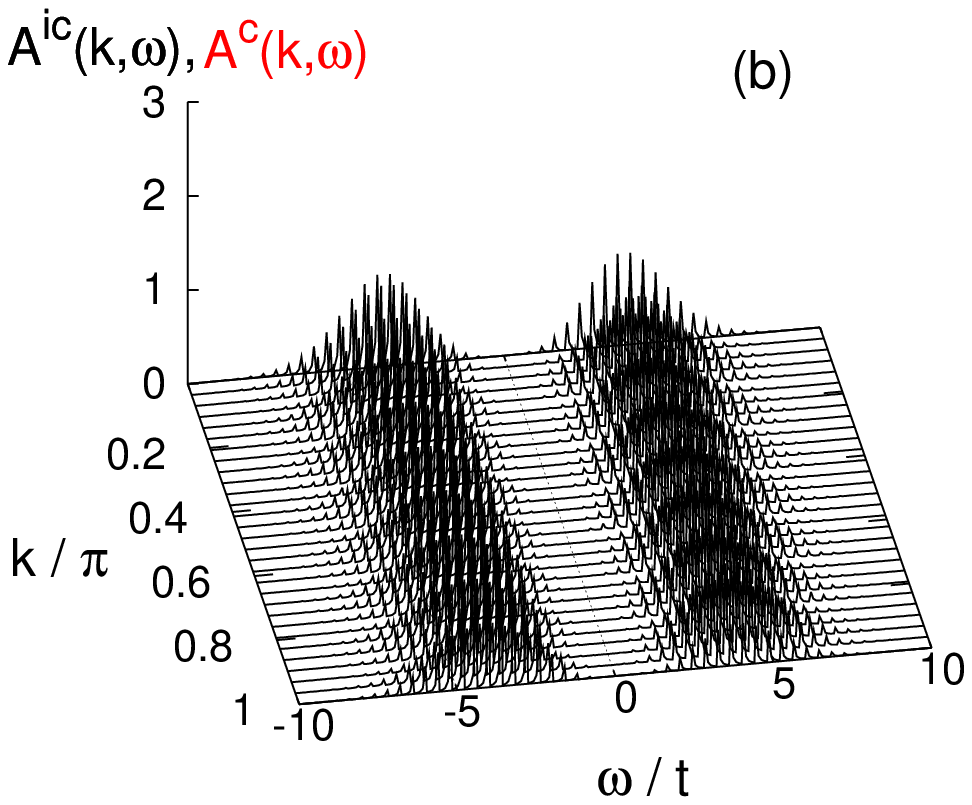}\\
  \includegraphics[width=0.495\textwidth]{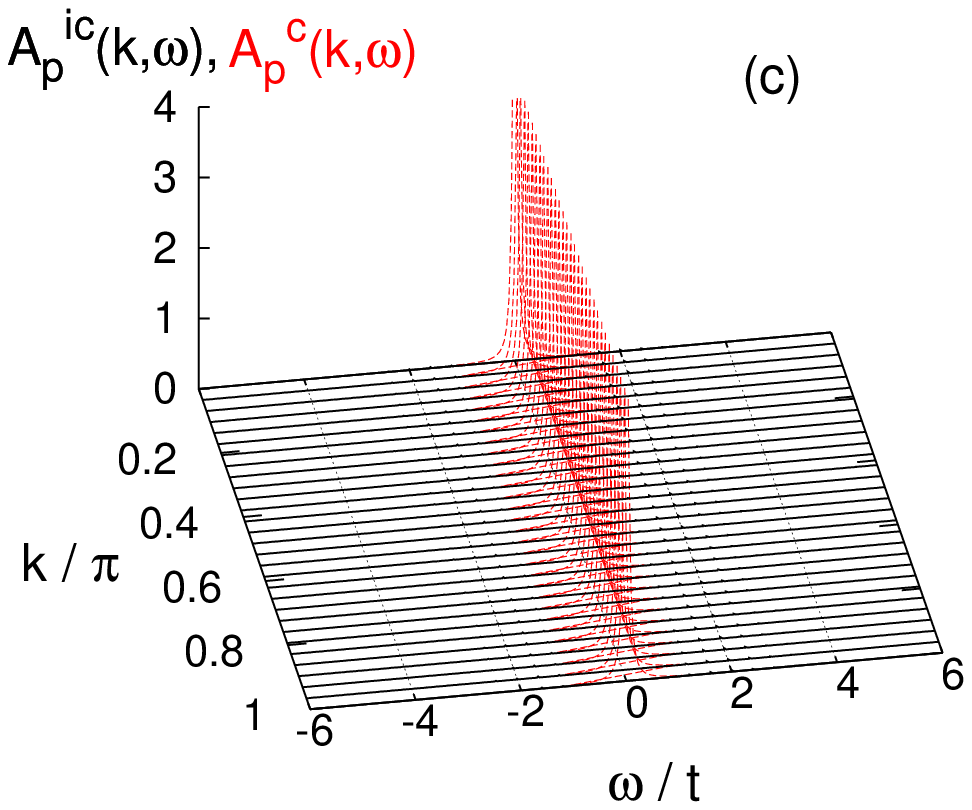}
  \end{center}
  \caption{\label{fig:cpt_an_akw_sc}%
    (colour online)
    As in figure~\ref{fig:cpt_an_akw_wc}, but for strong coupling
    $\lambda=2$. CPT results in (a) are for $N_\text{c}=8$ and $n=0.375$. Analytical
    results for the electronic spectrum (b) and the polaronic spectrum (c)
    are from the strong-coupling approximation and for $n=0.4$. ((b) and (c)
    taken from \cite{LoHoFe06})
  }
\end{figure}

\subsection{Strong coupling}\label{sec:sc}

At strong EP coupling, it is well known that the charge carriers are small
polarons at low carrier densities, and the spectral properties in this case
have been studied intensively (see, \eg, \cite{AlKaRa94,Stephan,HoAivdL03}).
The density dependence in the many-polaron case has been investigated in
\cite{HoNevdLWeLoFe04,LoHoFe06}, suggesting that virtually independent small
polarons exist even at large $n$ in the strong-coupling case.

CPT results for $\lambda=2$ and $n=0.375$ are shown in
figure~\ref{fig:cpt_an_akw_sc}(a) and agree well with previous QMC data
\cite{HoNevdLWeLoFe04}. The spectrum is dominated by incoherent
(multi-phonon) excitations well below and above the Fermi level, which reveal
a Poisson-like distribution and are centred close to $\Ep=4t$. No coherent
contributions are visible on the scale of the figure. Despite the
higher energy resolution of CPT as compared to QMC, we are not able to
monitor the coherent small-polaron band crossing the Fermi level,
which gives rise to a finite but small band conductivity. Calculations on
even larger clusters and including more Chebyshev moments are beyond our
present computational possibilities.

As in the weak-coupling case, an accurate picture of the physics can be
obtained from the analytical approach worked out in \cite{LoHoFe06}.
Figure~\ref{fig:cpt_an_akw_sc}(b) shows the electronic spectral function for
$n=0.4$ and the same parameters as in (a). The strong-coupling approximation
yields a coherent band with exponentially small spectral weight, and
the system may be well characterised as a polaronic metal. 

Further evidence for the polaronic nature of the quasiparticles is given by
the polaronic spectral function shown in figure~\ref{fig:cpt_an_akw_sc}(c).
Here only the coherent small-polaron 
band with spectral weight $z_k\approx 1$ is
visible, and any incoherent peaks---corresponding to electronic
contributions---are completely suppressed \cite{LoHoFe06}.  Due to the
approximations made, the density dependence of the analytical spectra comes
out too weak \cite{LoHoFe06} (as compared to QMC \cite{HoNevdLWeLoFe04}).

\subsection{Intermediate coupling}\label{sec:ic}

It has been stressed above that the intermediate-coupling regime is the most
interesting due to the existence of large polarons at low densities, whose
overlap gives rise to significant changes with increasing $n$. We chose
$\lambda=1$, \ie, close to the small-polaron crossover.

\begin{figure}
  \begin{center}
  \includegraphics[width=0.535\textwidth]{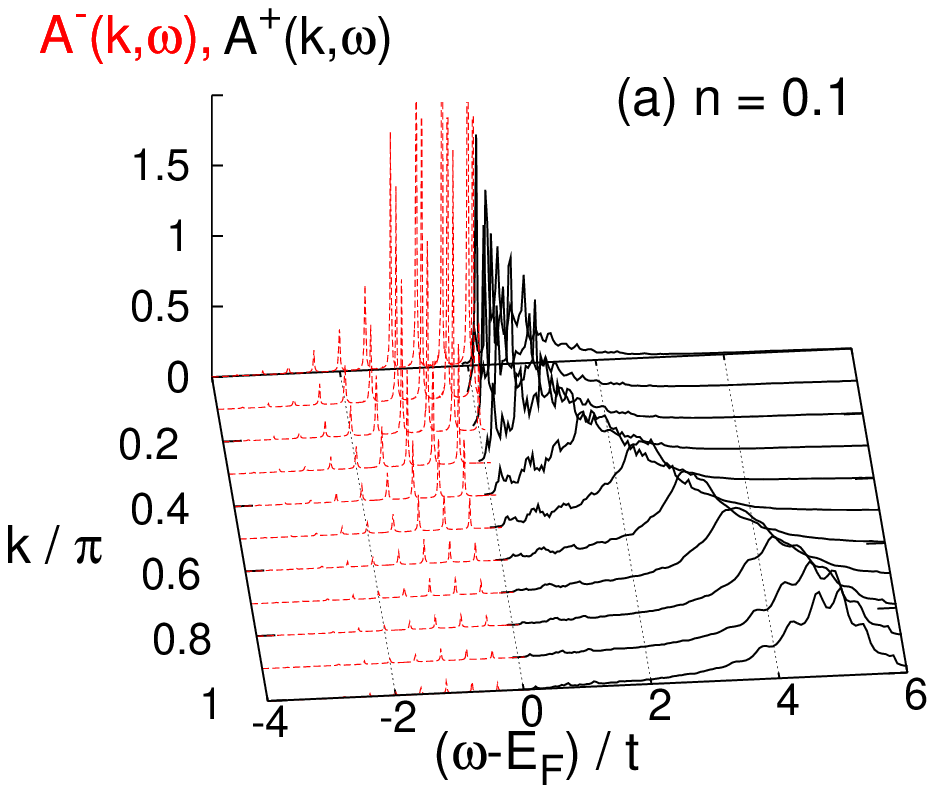}
  \includegraphics[width=0.455\textwidth]{bqw_n0.1.eps}\\
  \includegraphics[width=0.535\textwidth]{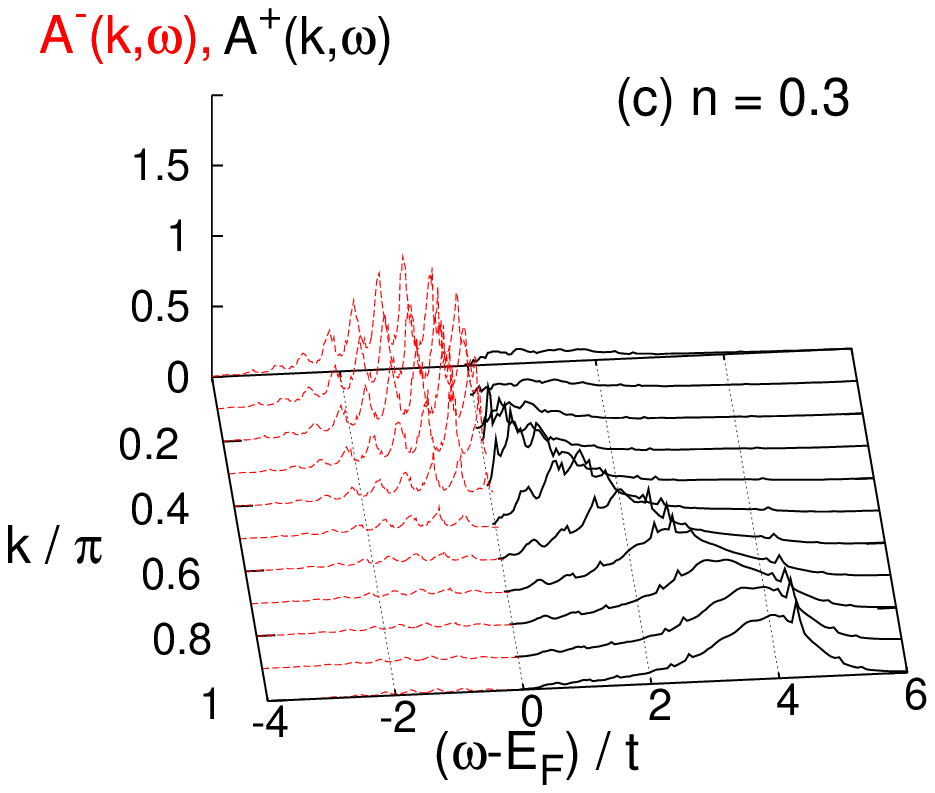}
  \includegraphics[width=0.455\textwidth]{bqw_n0.3.eps}
  \end{center}
  \caption{\label{fig:cpt_akw_ic}%
    (colour online) 
    (a) and (c): CPT results ($N_\text{c}=10$) for the single-particle spectral functions
    $A^-(k,\omega)$ ($\broken$, red) and $A^+(k,\omega)$ ($\full$,
    black) for $\gamma=0.4$, intermediate coupling $\lambda=1.0$ and
    different densities $n$. (taken from \cite{HoWeAlFe05})
    (b) and (d): phonon spectral function $B(q,\omega)$ ($\full$) from the cluster
    expansion ($N_\text{c}=10$) for the same parameters. Also shown are the
    polaron band dispersion  $E(k)-E(0)$ \cite{LoHoAlFe06} ($\chain$, red, in (b) only) and the bare
    phonon frequency ($\broken$). ((b) taken from \cite{LoHoAlFe06})}
\end{figure}

\paragraph{Photoemission and phonon spectra}
Figure~\ref{fig:cpt_akw_ic} shows the photoemission spectra for $n=0.1$ and 
0.3 (left panel), as well as the corresponding phonon spectra (right
panel). The results have been obtained by means of CPT and a cluster
expansion of the phonon self energy, respectively. Note that in the case of
CPT we have not analytically separated the symmetric $q=0$ phonon mode in
calculating $B(q,\om)$ \cite{WeWeAlScFe05}. This leads to increased numerical
effort for the same truncation error.

At low carrier density, figure~\ref{fig:cpt_akw_ic}(a), the
photoemission spectrum features a polaron band crossing $\Ef$, which flattens
at large $k$ where the excitation becomes phononic. Below $\Ef$, we see
non-dispersive peaks separated by $\om_0$ reflecting the Poisson distribution
of the phonons in the ground state. Above $\Ef$, incoherent
(phonon-mediated electronic) excitations form a broad band with a cosine-like
dispersion and a large width of about $4t$. Most importantly, this polaronic
spectrum is characterised by a separation between the coherent and incoherent
parts of the spectrum, \ie, no low-energy incoherent excitations exist.

The polaronic nature of the spectrum is also reflected in the phonon spectrum
$B(q,\om)$ (figure~\ref{fig:cpt_akw_ic}(b)). We see a clear signature
of the polaron band at low energies, which fits well the renormalised polaron
band dispersion in the thermodynamic limit. At higher energies, close to the
bare phonon energy, an almost flat band---overlaid by an excited ``mirror
polaron band''---is found \cite{LoHoAlFe06}.

Increasing the density to $n=0.3$ (figure~\ref{fig:cpt_akw_ic}(c)), we
find that a polaron band can hardly be identified. There is no longer a clear
separation between coherent and incoherent parts. Instead, a broad (width
$\approx 6t$) main band, formed by the merged phonon peaks below $\Ef$ and by
incoherent electronic excitations above $\Ef$, crosses the Fermi level. Such
a spectrum is reminiscent of the metallic system with carriers renormalised
by diffusive phonon scattering.

This qualitative change of the nature of the quasiparticles is also reflected
in the phonon spectrum. As shown in the figure~\ref{fig:cpt_akw_ic}(d), there
is no longer a well-defined polaronic signature in $B(q,\om)$ (cf
figure~\ref{fig:cpt_akw_ic}(b)). Instead, we observe at larger $q$ an
excitation band extending over a broad range of $\om$ values. There is also a
strong suppression of the signal at $\om=\om_0$, indicating that diffusive
scattering dominates.

\begin{figure}
  \centering
  \includegraphics[width=1\textwidth]{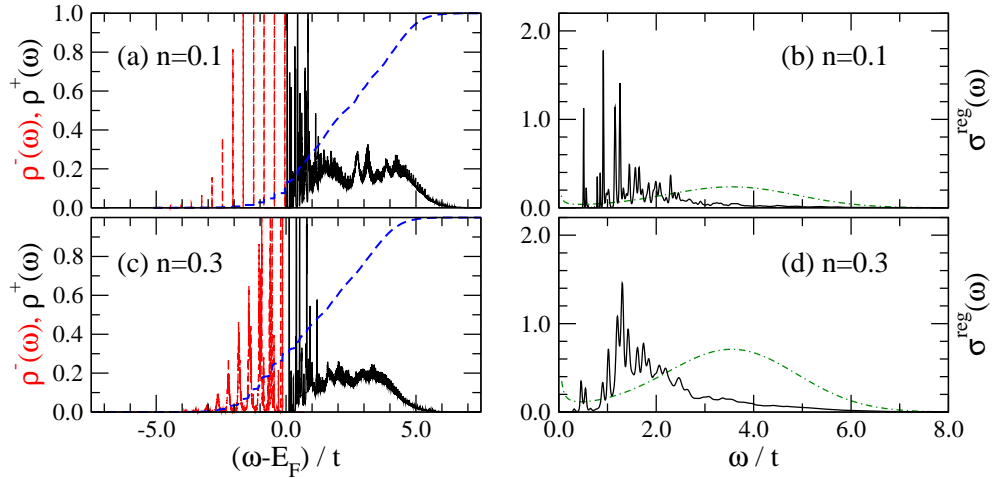}
  \caption{\label{fig:dos_oc}
    (colour online)
    (a) and (c): ED results ($N=10$) for the partial DOSs
    $\rho^{-}(\omega)$ ($\broken$, red) and $\rho^{+}(\omega)$ ($\full$, black)
    for $\lambda=1$, $\gamma=0.4$ and different band fillings $n$.
    (b) and (d):
    Regular part of the optical conductivity $\sigma^\text{reg}(\omega)$.
    Also shown ($\chain$, green): analytical strong-coupling result
    $\sigma^\text{reg} (\omega) = \sigma_0\,n\, (\omega_0g)^{-1}\,
    \omega^{-1}\, \exp [(\omega-2g^2\omega_0)/(2g\omega_0)]^{2}$
    ($\sigma_0=8$) \cite{Em93}. (taken from \cite{WeBiHoScFe05})
  }
\end{figure}

\paragraph{Optical response}
The density-driven changes are also visible in the optical response of the
system shown in figure~\ref{fig:dos_oc}. For the sake of clarity, we discuss
the latter together with results for the DOS which of
course reflects the features of the spectral function discussed above (cf
equation~(\ref{eq:dos})).

For $n=0.1$, we notice from the DOS a polaron feature at the Fermi level,
characterised by a (moderate) jump in the integrated DOS at $\Ef$ 
and the low spectral weight of $\rho^-(\omega<\Ef)$.  
Owing to the choice $\lambda=1$, \ie, the existence of a large polaron, 
the corresponding optical response $\sigma^\text{reg}(\omega)$ in
figure~\ref{fig:dos_oc}(b) strongly deviates from the analytical
strong-coupling result \cite{Em93}.  In particular, the maximum in
$\sigma^\text{reg}$ occurs well below the small-polaron value $2\Ep$.

At $n=0.3$ (figure~\ref{fig:dos_oc}(c) and (d)) the system shows enhanced transport.
The polarons are dissociated and the remaining electronic quasiparticles are
scattered by virtual phonons. As a consequence, a description of the optical
response in terms of small-polaron theory breaks down completely.

The change in the nature of the charge carriers in going from $n=0.1$ to 0.3
is also reflected in the Drude part $\mathcal{D}$ of the optical response.
The one-band sum rule for $\Re\sigma(\om)$ reads
$-E_\text{kin}/2=\mathcal{D}+\int_0^\infty\rmd\om \sigma^\text{reg}(\om)$.
Whereas for a single electron in the small-polaron regime the kinetic energy
is dominated by the regular part $\sigma^\text{reg}(\om)$ (\ie,
$\mathcal{D}\approx0$) \cite{LoHoAlFe07}, here we find an increase of the
ratio of $\mathcal{D}$ to $\int_0^\infty\rmd\om \sigma^\text{reg}(\om)$ from
3.9 ($n=0.1$) to 4.1 ($n=0.3$). This indicates that transport becomes more
coherent at large densities due to polaron dissociation.

\paragraph{Correlation functions}
\begin{figure}
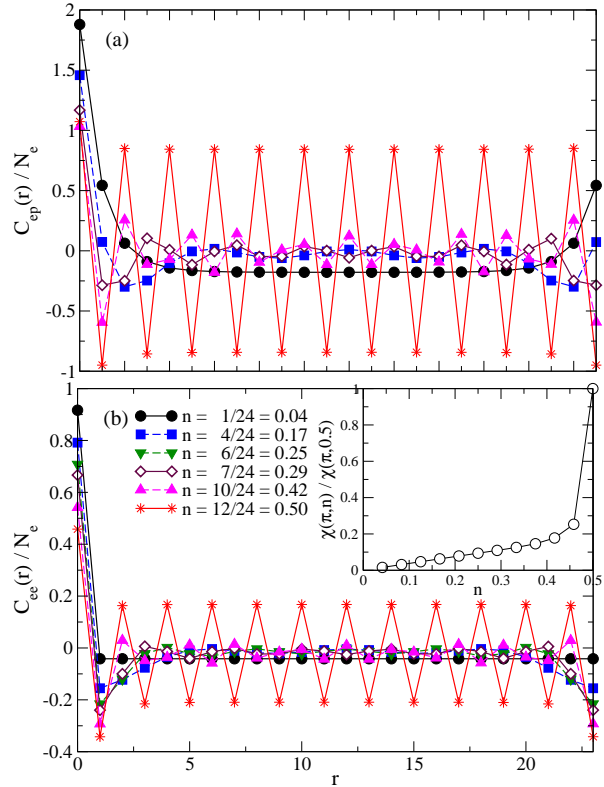

  \begin{center}
  \includegraphics[width=0.6\textwidth]{nx.eps}\\
  \includegraphics[width=0.6\textwidth]{nn.eps}
\end{center}
\caption{\label{fig:dmrg_corrfuncs}%
  (colour online) DMRG results ($N=24$) for the correlation functions (a)
  $C_\text{ep}(r)$ and (b) $C_\text{ee}(r)$ (normalised by the number of
  electrons $N_\text{e}$) for different band fillings $n$. The inset in (b)
  shows the renormalised charge-structure factor
  $\chi(\pi)=N^{-2}\sum_{ij} (-1)^{i+j}\las n_i n_j\ras$.
  Here, $\gamma=0.4$ and $\lambda=1$. Results have been obtained using
  anti-periodic (periodic) boundary conditions for even (odd) $N_\text{e}$
  (also in figure~\ref{fig:dmrg_energy}, see \cite{WeFe98}).} 
\end{figure}

The spectral properties calculated so far clearly show the qualitative change
in the nature of the ground state with increasing density. Since we
expect these changes to result from a dissociation of individual polarons, it
is highly desirable to calculate quantities which provide direct proof for
this mechanism. Therefore, we study here the correlation functions
$C_\text{ep}(r)$ and $C_\text{ee}(r)$ (see equations~(\ref{eq:Cep})
and~(\ref{eq:Cee})), which have been defined such as to permit comparison of
different $n$. Accounting for the homogeneous lattice distortion $\las
x\ras=2ng$ and the average electron density $n$, respectively, we have
$\sum_r C_\text{ep}(r)=0$ and $\sum_r C_\text{ee}(r)=0$. Furthermore, the
data in figure~\ref{fig:dmrg_corrfuncs} have been rescaled by the number of
electrons $N_\text{e}$.

Figure~\ref{fig:dmrg_corrfuncs}(a) shows $C_\text{ep}(r)$ for selected
densities $n$, obtained on a cluster with $N=24$ using the DMRG.
For the observables and parameters considered here, the results are
only weakly sensitive to the choice of boundary conditions (periodic or
anti-periodic, see caption of figure~\ref{fig:dmrg_corrfuncs}).

For a single electron ($n=0.04$), $C_\text{ep}(r)$ reveals the existence of a
large polaron with a lattice distortion extending over about five lattice
sites.  Already at $n=0.17$, the on-site ($r=0$) correlations are noticeably
reduced, whereas $C_\text{ep}(r>0)$ is increased as compared to $n=0.04$.
Increasing the density even further to $n=0.29$---roughly where a
polaron band can no longer be identified in the photoemission spectrum (see
figure~\ref{fig:cpt_akw_ic}(b))---we see a rather homogeneous value
$C_\text{ep}(r)=0$ (the average distortion has been subtracted) for almost
all values of $r$. This is exactly what we expect for a system of electrons
and unbound phonons.  $C_\text{ep}(r)$ starts to fluctuate as we go to even
large $n$ since we approach the Peierls transition. At $n=0.5$, in the
thermodynamic limit, the latter leads to long-range charge-density-wave order
with alternating occupied and empty lattice sites, causing symmetric
fluctuations in $C_\text{ep}(r)$ \cite{HoWeBiAlFe06}.

The corresponding results for the fermion-fermion correlation function
$C_\text{ee}(r)$ show a very similar density dependence. An interesting open
question concerns the possibility of charge density wave  
formation in the present model at
other commensurate densities such as $1/4$ or $1/3$. Although the critical EP
coupling for the transition to such insulating states is expected to be
significantly larger than at half filling, due to reduced Umklapp scattering
and the absence of perfect nesting, figure~\ref{fig:dmrg_corrfuncs} indeed
reveals $N_\text{e}$ maxima for, \eg, $n=0.25$. In the thermodynamic limit, 
charge ordering may thus be realised at commensurate fillings away from 
$n=0.5$.

\paragraph{Absence of phase separation}
Finally, we consider the possibility of phase separation of polarons.
It is known from the spinful Holstein model that real-space pairing of
electrons into bipolarons can occur if the EP-mediated attraction overcomes
the kinetic and Coulomb energy. Hence, one might expect aggregation of
polarons in one region of the system, \ie, a clustering of polarons (polaron
droplets) or bipolarons. Phase separation might also appear when doping the 
system away from the half-filled (charge-ordered) band case. 

\begin{figure}
  \begin{center}
  \includegraphics[width=0.5\textwidth]{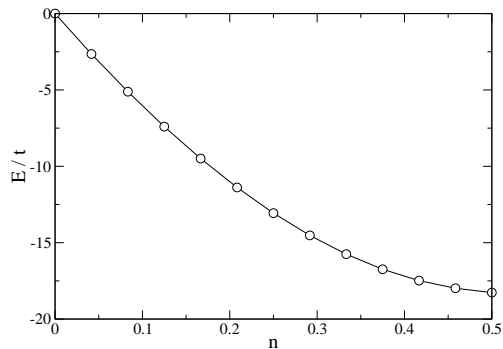}\\
\end{center}
\caption{\label{fig:dmrg_energy}%
    DMRG results for the ground-state energy $E$ as a
    function of band filling $n$. Here $N=24$, $\gamma=0.4$ and
    $\lambda=1$. Lines are guides to the eye.}
\end{figure}

To approve or rule out this possibility, we calculate the ground-state energy 
as a function of carrier density using the DMRG. 
The latter approach permits us to consider a large cluster with
$N=24$---the number of fermions $N_\text{e}=1\dots12$---so that we
can increment $n$ in rather small steps. Any tendency toward phase separation
should manifest itself by non-convex behaviour of $E$ as a function of $n$.
The results, presented in figure~\ref{fig:dmrg_energy}, provide strong
evidence for the absence of phase separation in the model considered here.
As a consequence of the very small binding energy in the model with two
electrons \cite{Hovdl2006}, the tendency towards pairing is also suppressed
as compared to the spinful case.

%

%
%
%
\section{Conclusions}\label{sec:conclusions}
%
%
%

Using a variety of different and in many aspects complementary methods, we
have obtained a rather complete understanding of the many-polaron
problem in the framework of the one-dimensional spinless Holstein model.
Whereas the physics is simple in the limiting cases of weak or strong
electron-phonon interaction, or in the non-adiabatic regime---the charge
carriers being either weakly renormalised electrons or small
polarons---substantial density effects are observed in the (adiabatic)
intermediate-coupling case.

Starting from low densities $n\ll1$ (the single-polaron problem), the nature
of the charge carriers changes from large polarons to renormalised electrons,
resulting in a metallic system at intermediate carrier densities
$n\approx0.3$. This crossover has been investigated by studying
photoemission spectra, phonon spectra, optical response, fermion-lattice and
fermion-fermion correlation functions. All these observables support the
hypothesis of dissociation of large polarons in a conclusive way.

Furthermore, by calculating the ground-state energy as a function of $n$ up
to $n=0.5$, we can strongly argue against phase separation of polarons.
Interestingly, the results point toward the occurrence of lattice 
instabilities accompanied by charge ordering at 
commensurate fillings other than 0.5, but this issue needs
further investigation.

The present work is restricted to a rather simple model. However, the density
effects on the nature of the charge carriers arise from the residual
interaction (overlap) of extended polarons. Such a basic mechanism may be
expected to be important also in more involved models (\eg, with long-range
interactions and spin, or in higher dimensions---large polarons exist in
Fr\"ohlich models also for $\rD>1$).

There is also no reason to expect the absence of similar physics in dense
polaronic systems such as the manganites, especially as recent experimental
work points towards the existence of large polarons in these materials
\cite{HaMaDeLoKo04}. The relevance of the effects discussed in this work to
realistic materials is further substantiated by the fact that intermediate
densities, small but finite phonon frequencies and intermediate couplings are
widely regarded as the experimentally most relevant parameter region.

Finally, since existing weak-coupling or low-density theories on the same or
similar models do not exhibit the crossover considered here due to the
neglect (or insufficient treatment) of density effects, future work on more
general many-polaron systems is highly desirable. Despite the greater
complexity, it would be important to take into account the spin degrees of
freedom, as well as a finite Coulomb repulsion between charge carriers. The
rapid increase in computer power opens up the perspective of such studies in
the near future.

\ack

We gratefully acknowledge financial support by the Austrian Science Fund
(FWF) through the Erwin-Schr\"odinger Grant No~J2583, the Deutsche
Forschungsgemeinschaft through SPP1073, the DFG, KONWIHR, and the European
Science Foundation. We would like to thank A~Alvermann and J~Loos 
for valuable discussion.



\section*{References}


\end{document}